\newcommand{\cl}{\centerline}
\renewcommand{\theequation}{\thesection.\arbic{equation}}
\renewcommand{\theequation}{A.\arbic{equation}}

\newcommand\beq{\begin{equation}}
\newcommand\eeq{\end{equation}}
\newcommand\bea{\begin{eqnarray}}
\newcommand\eea{\end{eqnarray}}
\documentstyle[12pt]{article}
\begin{document}
\begin{titlepage}
\setlength{\textwidth}{5.0in}
\setlength{\textheight}{7.5in}
\setlength{\parskip}{0.0in}
\setlength{\baselineskip}{18.2pt}
\hfill
SNUTP97/007
\vfill
\cl{\Large{{\bf $\eta$-$\eta^{\prime}$ Mixture in Chiral Bag}}}\par
\vskip 8.mm
\cl{Soon-Tae Hong, Dong-Pil Min}
\vskip 0.4cm
\cl{Department of Physics and Center for Theoretical Physics,}\par
\cl{Seoul National University}
\cl{Seoul 151-742, Korea}
\vskip 1.0cm
\cl{\today}
\vskip 1.5cm
\vfill
\begin{center}
{\bf ABSTRACT}
\end{center}
\begin{quote}
The $\eta$-$\eta^{\prime}$ mixture is discussed in the chiral bag model to
calculate the pseudoscalar octet-singlet mixing angle consistent with the
experimental data.  The color anomaly is taken into account with the modified
boundary conditions, which shows the relation between the $\eta^{\prime}$
mass and the gluon condensate inside the chiral bag.  We show, however, that
$\eta$-$\eta^{\prime}$ mixing angle can follow the Cheshire Cat Principle,
i.e., insensitivity to the bag radius. 
\end{quote}
\end{titlepage}
\par
 
\section{Introduction}\par
\setcounter{equation}{0}
\renewcommand{\theequation}{\arabic{section}.\arabic{equation}}

Nearly all known mesons can be understood as bound states of a quark q 
and antiquark $\bar{\rm q}$ (the flavors of q and 
$\bar{\rm q}$ may be different).  The nine possible $\bar{\rm q}$q combinations
containing u, d and s quarks group themselves into an octet and
a singlet:
$$
{\bf 3}\otimes\bar{\bf 3}={\bf 8}\oplus{\bf 1}.
$$
States with the same $IJ^{P}$ and additive quantum numbers can mix (if they
are eigenstates of charge conjugation $C$, they must also have the same value
of $C$).  Thus the $I=0$ member of the ground state pseudoscalar octet mixes
with the corresponding pseudoscalar singlet to yield the $\eta$ and
$\eta^{\prime}$.  These appear as members of a nonet.

For the pseudoscalar mesons the Gell-Mann-Okubo formula is given by
\beq
m_{\eta}^{2}=\frac{4}{3}m_{K}^{2}-\frac{1}{3}m_{\pi}^{2}\label{gmo}
\eeq
assuming no octet-singlet mixing, which is extremely sensitive to SU(3)
symmetry breaking.

The value of the $\eta$-$\eta^{\prime}$ mixing angle has been the subject of
discussion almost from the time that SU(3) flavor symmetry was proposed.
In the simplest possible situation where one assumes the presence of only an 
octet and a singlet, the quadratic Gell-Mann-Okubo mass formular yields a 
pseudoscalar mixing angle of $\theta=-10^{\rm o}$.  With the same assumption a 
Gell-Mann-Okubo mass formular which is linear in the masses  gives $\theta=
-23^{\rm o}$.  For reasons that have to do with both theory and experiment at
a given time over the years most authors\cite{author,author2} have taken
$\theta= -10^{\rm o}$.

Nowadays there has been significant discussion concerning the strangeness in 
the nucleon structure.  Especially the measurement of the spin structure
function of the proton given by European Muon Collaboration (EMC) experiment on
deep inelastic muon scattering\cite{emcexp} has suggested a lingering question
touched on by physicists that the effect of strange quarks on nucleon
structure is not small.  The EMC result has been interpreted as the possibility
of a strange quark sea strongly polarized opposite to the proton spin. 
Similarly such interpretation of the strangeness has been brought to other
analyses of low energy elastic neutrino-proton scattering\cite{scatt} and the
kaon condensation\cite{cond,cond2} in the neutron star matter.

On the other hand, the chiral bag model (CBM)\cite{cbm} couples fundamental
hadron constituents inside to the pseudoscalar meson fields obeying nonlinear
chiral lagrangian outside the chiral bag through the boundary term on the bag
surface introduced to restore the chiral invariance.

The CBM has enjoyed considerable success in predictions\cite{pred,pred2} of the
baryon static properties such as the EMC experiments and the magnetic moments
of baryon octet.  After the discovery of the Cheshire Cat Principle
\cite{ccp} the CBM has been also regarded as a candidate which unifies the MIT 
bag and Skyrme models and gives model-independent relations insensitive to the 
bag radius.  Morover Brown et al.\cite{brown} have calculated the pion-cloud 
contributions to the bayon magnetic moments by using the SU(2) CBM as an 
effetive nonrelativistic quark model.  This scheme has been generalized
\cite{nrqm,nrqm2} to SU(3) CBM to yield the minimal multi-quark structure
\cite{min} so that the meson-cloud could be generated inside the chiral bag in
terms of the nonperturbative higher representation mixing in the wave functions of the baryons. 

In this paper we will introduce the SU(3) symmetry breaking terms to produce 
the $\eta$-$\eta^{\prime}$ mixture and to estimate the octet-singlet mixing
angle, under the assumption that the physical states are orthogonal, namely,
the mixing is independent of energy.  We will also investigate the color
anomaly in terms of the gluon condensate and $\eta^{\prime}$ mass.  The quark
condensate will be discussed to yield the pseudoscalar octet masses.  Then we
will show that the $\eta$-$\eta^{\prime}$ mixing angle is insensitive to the
bag radius in accordance with the Cheshire Cat Principle.

In Section 2, the CBM lagrangian with extended boundary condition will be
introduced. 

In Section 3, the $\eta$ degrees of freedom will be discussed both inside and
outside the chiral bag to incorporate the $\eta$-$\eta^{\prime}$ mixing, so
that one can have nontrivial FSAC and estimate the pseudoscalar mixing angle. 
The color anomaly will be also discussed together with the gluon and quark
condensates.  Our conclusions will be found in Section 4.

\section{Model with Extended Boundary Condition}\par 
\setcounter{equation}{0} 
\renewcommand{\theequation}{\arabic{section}.\arabic{equation}} 

Now in order to introduce the $\eta$-$\eta^{\prime}$ mixing effects in the CBM
we start with the lagrangian of the form 
\begin{eqnarray} 
& &{\cal L}={\cal L}_{QCD}+{\cal L}_{M}(={\cal L}_{CS}+{\cal L}_{CSB}
	   +{\cal L}_{FSB})+{\cal L}_{I}\label{lag}\\ 
& &{\cal L}_{QCD}=(\bar{\psi}i\gamma^{\mu}D_{\mu}\psi
       -\bar{\psi}M\psi-\frac{1}{2}{\rm tr} F_{\mu\nu}F^{\mu\nu})\Theta_{B}
	\label{qcd}\\
& &{\cal L}_{CS}=(-\frac{1}{4}f_{\pi}{\rm tr}(l_{\mu}l^{\mu})+\frac{1}{32e^{2}}
       [l_{\mu},l_{\nu}]^{2}+{\cal L}_{WZW}) \bar{\Theta}_{B}\label{lagcs}\\ 
& &{\cal L}_{CSB}=\frac{1}{4}f_{\pi}^{2}m_{\pi}^{2}({\rm tr}(U+U^{\dagger}
     -2)-\frac{1}{3}\epsilon(-i{\rm ln}{\rm det}U)^{2})\bar{\Theta}_{B}
		      \label{lagcsb}\\
& &{\cal L}_{FSB}=\frac{1}{6}f_{\pi}^{2}(\chi^{2}m_{K}^{2}-m_{\pi}^{2})
   {\rm tr}((1-\sqrt{3}\lambda_{8})(U+U^{\dagger}-2))\bar{\Theta}_{B}
	 \nonumber\\
   & &\ \ \ \ \ \ \ \ \ \   -\frac{1}{12}f_{\pi}^{2}(\chi^{2}-1){\rm tr}
      ((1-\sqrt{3}\lambda_{8})
      (Ul_{\mu}l^{\mu}+l_{\mu}l^{\mu}U^{\dagger}))\bar{\Theta}_{B}
      \label{lagfsb}\\
& &{\cal L}_{I}=\frac{1}{2}\bar{\psi}U_{5}\psi\Delta_{B}
\end{eqnarray}
where one has the quark field $\psi$ with SU(3) degrees of freedom and the 
gluon field strength tensor $F_{\mu\nu}=\partial_{\mu}G_{\nu}-\partial_{\nu}
G_{\mu}+ig[G_{\mu},G_{\nu}]$ inside the bag.  Here $D_{\mu} 
=\partial_{\mu}+igG_{\mu}^{a}\frac{\lambda_{a}}{2}$ is the covariant 
derivative with the effective strong coupling constant $g$. 
And Gell-Mann matrices $\lambda_{a}$ are normalized to satisfy $\lambda_{a}
\lambda_{b}=\frac{2}{3}\delta_{ab}+(if_{abc}+d_{abc})\lambda_{c}$ and
$\Theta_{B}(=1-\bar{\Theta}_{B})$ is the bag theta function (one inside the 
bag and zero outside the bag).  

In the ref.\cite{nrqm} the authors have considered the case SU(3)$\times$SU(3)
chiral theory without $\eta$ degrees of freedom in the Skyrme phase. Also it
was suggested that inside the bag the quark-antiquark annihilation channel
could be induced via anomalous gluon effect in the U(1) flavor sector.  Now we
extend to the chiral bag with U(3)$\times$U(3) group structure so that we can
incorporate gluons and the $\eta$ fields consistently.  To do this we modify
the chiral field $U$ as follows
\beq
U={\rm exp}(i(\lambda_{0}\pi_{0}+\lambda_{a}\pi_{a})/f_{\pi}),\ \ \ \ 
U_{5}={\rm exp}(i\gamma_{5}(\lambda_{0}\pi_{0}+\lambda_{a}\pi_{a})/f_{\pi})
\eeq
where $\pi_{0}$ is the zeroth eta field and $\lambda_{0}=\sqrt{2/3}I (I:
3\times 3$ unit matrix).  Thus outside the bag the chiral field $U$ is
described by the pseudoscalar meson fields $\pi_{0}$ and $\pi_{a} (a=1,...,8)$. Here $l_{\mu}=U^{\dagger}\partial_{\mu}U$ and ${\cal L}_{WZW}$ stands for the
topological Wess-Zumino-Witten term.  In the numerical calculation we will use
the parameter fixing e=4.75, $f_{\pi}=93$ MeV and $f_{K}=114$ MeV.  The surface
interaction term ${\cal L}_{I}$ typical in the CBM plays crucial role in the
restoration of the chiral symmetry by coupling the pseudoscalar meson fields to
the quarks on the bag boundary.

Here one notes that, together with the gluon and quark mass terms in
(\ref{qcd}), the terms in ${\cal L}_{CSB}$ break the chiral symmetry
and the pion mass term in ${\cal L}_{CSB}$ is chosen such that it will vanish
for $U=1$.  Also the SU(3) flavor symmetry breaking with $m_{K}/m_{\pi}\neq 1$
and $\chi=f_{K}/f_{\pi}\neq 1$ is included in ${\cal L}_{FSB}$.
             
Even though the mass terms in ${\cal L}_{QCD}$, ${\cal L}_{CSB}$ and 
${\cal L}_{FSB}$ break both the SU(3)$\times$SU(3) and the diagonal SU(3)
symmetry so that chiral symmetry cannot be conserved, these terms without
derivatives yield no explicit contribution on the flavor singlet axial
currents (FSAC) and at least in the adjoint representation of the SU(3) group
the FSAC are conserved and  of the same form as the chiral limit result.  
However the kinetic term in ${\cal L}_{FSB}$ gives rise to the chiral symmetry
broken FSAC.

\section{$\eta$-$\eta^{\prime}$ Mixture}\par 
\setcounter{equation}{0} 
\renewcommand{\theequation}{\arabic{section}.\arabic{equation}} 

In the chiral symmetric case we have seen that the FSAC vanishes at leading
order in $1/N_{c}$ since for example the $\pi_{0}$ decouples from the other
mesons in the Skyrmion limit, namely $g_{\pi_{0}NN}=0$.  In order to obtain
nontrivial contribution to the FSAC we need to include the
$\eta$-$\eta^{\prime}$ mixing by introducing to the lagrangian (\ref{lag}) the
symmetry breaking terms ${\cal L}_{CSB}+{\cal L}_{FSB}$ as mentioned above.

Adding the kinetic term in ${\cal L}_{CS}$ in (\ref{lagcs}) to
${\cal L}_{FSB}$ in (\ref{lagfsb}) we obtain the bilinear kinetic terms in the
weak field approximation
\beq
{\cal L}_{kin}=\frac{1}{2}\partial_{\mu}\pi_{i}\partial^{\mu}\pi_{i}
               +\frac{1}{2}\partial_{\mu}\pi_{M}\partial^{\mu}\pi_{M}
               +\frac{1}{2}\partial_{\mu}\tilde{\pi}^{\dagger}{\cal M}
	        \partial^{\mu}\tilde{\pi}
\label{lagkin}
\eeq
where $\pi_{i}$ ($i=1,2,3$) and $\pi_{M}$ ($M=4,5,6,7$) are the pion and kaon
fields and $\tilde{\pi}^{\dagger}=(\pi_{0},\pi_{8})$.  The matrix elements of 
${\cal M}$ are then given by
\bea
{\cal M}_{11}&=& (1+2\delta)\frac{f_{\pi}^{2}}{f_{\pi_{0}}^{2}}\nonumber\\
{\cal M}_{22}&=& (1+4\delta)\frac{f_{\pi}^{2}}{f_{\pi_{8}}^{2}}\nonumber\\
{\cal M}_{12}&=& {\cal M}_{21}=-2\sqrt{2}\delta\frac{f_{\pi}^{2}}
		 {f_{\pi_{0}}f_{\pi_{8}}}
\eea
where $\delta=\frac{1}{3}(\chi^{2}-1)$.  In the $\pi_{0}$-$\pi_{8}$ channel we
diagonalize the matrix ${\cal M}$ with the physical fields defined as
\bea
\eta&=&(1+2\delta)\frac{f_{\pi}}{f_{\pi_{8}}}\pi_{8}
       +a\delta\frac{f_{\pi}}{f_{\pi_{0}}}\pi_{0}
       =\pi_{8}\cos\theta-\pi_{0}\sin\theta\nonumber\\
\eta^{\prime}&=&b\delta\frac{f_{\pi}}{f_{\pi_{8}}}\pi_{8}
       +(1+\delta)\frac{f_{\pi}}{f_{\pi_{0}}}\pi_{0}
       =\pi_{8}\sin\theta+\pi_{0}\cos\theta\label{eta}\label{trf}
\eea
where $\theta$ is the $\eta$-$\eta^{\prime}$ mixing angle.

Using the above eta fields (\ref{eta}) we can obtain the desired quadratic
kinetic terms as follows
\bea
\frac{1}{2}\partial_{\mu}\eta\partial^{\mu}\eta
+\frac{1}{2}\partial_{\mu}\eta^{\prime}\partial^{\mu}\eta^{\prime}
&=&\frac{1}{2}(1+4\delta)\frac{f_{\pi}^{2}}{f_{\pi_{8}}^{2}}
\partial_{\mu}\pi_{8}\partial^{\mu}\pi_{8}
+\frac{1}{2}(1+2\delta)\frac{f_{\pi}^{2}}{f_{\pi_{0}}^{2}}
\partial_{\mu}\pi_{0}\partial^{\mu}\pi_{0}\nonumber\\
& &+(a+b)\delta\frac{f_{\pi}^{2}}{f_{\pi_{0}}f_{\pi_{8}}}
\partial_{\mu}\pi_{8}\partial^{\mu}\pi_{0}\label{etakin}
\eea
to yield the relations
\bea
f_{\pi_{8}}^{2}&=&(1+4\delta)f_{\pi}^{2}\cong f_{\eta}^{2}\label{rel11}\\
f_{\pi_{0}}^{2}&=&(1+2\delta)f_{\pi}^{2}\cong f_{\eta^{\prime}}^{2}
		  \label{rel12}\\
a+b&=&-2\sqrt{2}\label{rel13}.
\eea
Here one notes that there is a residual ambiguity in the choice of $(a,b)$
which corresponds to the freedom of performing an orthogonal transformation
on (\ref{etakin}).  That ambiguity can be removed by considering the mass
quadratic terms
\beq
{\cal L}_{mass}=-\frac{1}{2}m_{\pi}^{2}\pi_{i}^{2}
		-\frac{1}{2}m_{K}^{2}\pi_{M}^{2}
                -\frac{1}{2}m_{\pi}^{2}\tilde{\pi}^{\dagger}{\cal N}\tilde{\pi}
\label{lagmass}
\eeq
where the mass matrix elements are given by
\bea
{\cal N}_{11}&=&\frac{2}{3}m_{K}^{2}\frac{f_{K}^{2}}{f_{\pi_{0}}^{2}}
   +(\frac{1}{3}+\epsilon)m_{\pi}^{2}\frac{f_{\pi}^{2}}{f_{\pi_{0}}^{2}}
   \nonumber\\
{\cal N}_{22}&=&\frac{4}{3}m_{K}^{2}\frac{f_{K}^{2}}{f_{\pi_{8}}^{2}}
   -\frac{1}{3}m_{\pi}^{2}\frac{f_{\pi}^{2}}{f_{\pi_{8}}^{2}}
   \nonumber\\
{\cal N}_{12}&=&{\cal N}_{21}=-\frac{2\sqrt{2}}{3}(m_{K}^{2}\frac{f_{K}^{2}}
   {f_{\pi_{0}}f_{\pi_{8}}}
   -m_{\pi}^{2}\frac{f_{\pi}^{2}}{f_{\pi_{0}}f_{\pi_{8}}}).
\eea
Again we diagonalize the matrix ${\cal N}$ with the transformation (\ref{trf})
to yield the quadratic mass terms
\bea
-\frac{1}{2}m_{\eta}^{2}\eta^{2}
-\frac{1}{2}m_{\eta^{\prime}}^{2}\eta^{\prime 2}
&=&-\frac{1}{2}m_{\eta}^{2}
    ((1+4\delta)\frac{f_{\pi}^{2}}{f_{\pi_{8}}^{2}}\pi_{8}^{2}
   +2a\delta\frac{f_{\pi}^{2}}{f_{\pi_{0}}f_{\pi_{8}}}\pi_{8}\pi_{0})
    \nonumber\\
& &-\frac{1}{2}m_{\eta^{\prime}}^{2}
   (2b\delta\frac{f_{\pi}^{2}}{f_{\pi_{0}}f_{\pi_{8}}}\pi_{0}\pi_{8}
    +(1+2\delta)\frac{f_{\pi}^{2}}{f_{\pi_{0}}^{2}}\pi_{0}^{2})
\eea  
so that we obtain the other relations
\bea
m_{\eta^{\prime}}^{2}&=&\frac{2}{3}\frac{1+3\delta}{1+2\delta}m_{K}^{2}
		      +\frac{1}{3}\frac{1+3\epsilon}{1+2\delta}m_{\pi}^{2}
		      \label{rel21}\\
m_{\eta}^{2}&=&\frac{4}{3}\frac{1+3\delta}{1+4\delta}m_{K}^{2}
		      -\frac{1}{3}\frac{1}{1+4\delta}m_{\pi}^{2}
		      \label{rel22}\\ 
am_{\eta}^{2}+bm_{\eta^{\prime}}^{2}&=&-\frac{2\sqrt{2}}{3\delta}
			 ((1+3\delta)m_{K}^{2}-m_{\pi}^{2})\label{rel23}.
\eea
Here one notes that in the limit $\delta=0$ where $f_{K}=f_{\pi}$ the relation
(\ref{rel22}) reproduces the Gell-Mann-Okubo formula (\ref{gmo}).  Also the
relation (\ref{rel21}) determines the constant $\epsilon$ as below
\beq
\epsilon=(1+2\delta)\frac{m_{\eta^{\prime}}^{2}}{m_{\pi}^{2}}
	 -\frac{2}{3}(1+3\delta)\frac{m_{K}^{2}}{m_{\pi}^{2}}-\frac{1}{3}
\eeq
and the combination of (\ref{rel13}) and (\ref{rel23}) yields
\beq
a=\frac{1}{m_{\eta^{\prime}}^{2}-m_{\eta}^{2}}\frac{2\sqrt{2}}{3\delta}
    ((1+3\delta)m_{K}^{2}-m_{\pi}^{2}-3\delta m_{\eta^{\prime}}^{2}).
\label{a}
\eeq

Now we use the experimental data\cite{exp} $f_{K}=114$ MeV, $m_{\pi}=140$ MeV,
$m_{K}=496$ MeV, $m_{\eta^{\prime}}=960$ MeV and $m_{\eta}=550$ MeV to estimate
the above parameters as follows
$$
a=-1.037,\ \ \ b=-1.790,\ \ \ \delta=0.168
$$
and the $\eta$-$\eta^{\prime}$ mixing angle 
\beq
\theta=-12.7^{\rm o}
\eeq
which is qualitatively consistent with phenomenological analyses\cite{theta}.
	
On the other hand inside the bag surface we need to have the mechanism 
consistent with the pseudoscalar octet-singlet mixing in the Skyrme phase if we
assume that the Cheshire Cat Principle holds in the CBM.  In ref.\cite{nrqm} it has been shown that the inside-mesons orginate from the minimal multi-quark
Fock space qqq+qqq$\bar{\rm q}$q whose possible SU(3) representations are
constrained by the Clebsch-Gordan series ${\bf 8}\oplus\bar{\bf 10}\oplus
{\bf 27}$.  Thus one could have the meson-cloud, namely $\bar{\rm q}$q content, inside the bag through the channel of qqq$\bar{\rm q}$q multi-quark Fock space.
Here the mesonic $\bar{\rm q}$q contents refer to all the possible flavor
combinations to construct the pseudoscalar mesons inside the bag.

Until now there is no explicit known mechanism to explain the meson cloud
inside the chiral bag.  Presumably the mechanism seems closely related to the
pseudoscalar composite operators $\bar{\psi}i\gamma_{5}\lambda_{0}\psi
\sim\pi_{0}$ and $\bar{\psi}i\gamma_{5}\lambda_{a}\psi\sim\pi_{a}$
$(a=1,...,8)$ and to the quark-antiquark annihilation channel\cite{ann} through
the anomalous gluon effect.  In the U(1) flavor sector the gluons are thus
supposed to mediate the pseudoscalar meson via the $\bar{\rm q}$q pair creation
and annihilation mechanism.  In this mechanism the composite operators
$\bar{\psi}i\gamma_{5}\lambda_{0}\psi$ and $\bar{\psi}i\gamma_{5}
\lambda_{8}\psi$ corresponding to $\pi_{0}$ and $\pi_{8}$ respectively could
mix to yield the pseudoscalar octet-singlet mixture 
\bea
\eta&\sim&\bar{\psi}i\gamma_{5}(\lambda_{8}\cos\theta-\lambda_{0}\sin\theta)\psi
\nonumber\\
\eta^{\prime}&\sim&\bar{\psi}i\gamma_{5}(\lambda_{8}\sin\theta+\lambda_{0}\cos
		\theta)\psi\nonumber
\eea
so that the Cheshire Cat Principle could be satisfied and one could have the 
same $\eta$-$\eta^{\prime}$ mixing angle both inside and outside the chiral
bag.

Moreover the presence of the $\eta^{\prime}$ field at the boundary induces a 
color electric field normal to the bag surface in the form $\eta^{\prime}
\hat{n}\cdot\vec{B}^{a}/f_{\eta^{\prime}}$.  As a result color will be pulled
out of the bag at a rate proportional to $\dot{\eta}^{\prime}\hat{n}\cdot
\vec{B}^{a}/f_{\eta^{\prime}}$.  This phenomenon is essentially the color
anomaly\cite{color,color2}.  At the quantum level a counter term thus has to
be added at the bag boundary which exactly cancels the induced color electric
field.  The most general form of the surface charge counter term is given
by\cite{color}
\beq
{\cal L}_{CT}=\frac{ig^{2}}{32\pi^{2}}\oint_{\Sigma}{\rm d}\beta K_{5}^{\mu}
	      n_{\mu}({\rm tr}{\rm ln}U^{\dagger}-{\rm tr}{\rm ln}U)
\eeq
where $\beta$ is a point on the bag surface $\Sigma$ and $K_{5}^{\mu}=\epsilon
^{\mu\nu\rho\sigma}(G_{\nu}^{a}F_{\rho\sigma}^{a}-\frac{2}{3}f^{abc}gG_{\nu}
^{a}G_{\rho}^{b}G_{\sigma}^{c})$ is the properly regularized Chern-Simons
current.  Also the counter term induces changes in the boundary conditions for
the color electric and magnetic fields in the quasi-abelian approximation as
below\footnote{Here one notes that in the U$_{A}(1)$ channel we have used
$\eta^{\prime}$ field instead of $\pi_{0}$, ignoring the possible
$\eta$-$\eta^{\prime}$ mixing which is the negligible secondary effect in the
color anomaly.}
\bea
\hat{n}\cdot\vec{E}^{a}&=&-\frac{N_{f}g^{2}}{8\pi^{2}f_{\eta^{\prime}}}
			   \hat{n}\cdot\vec{B}^{a}\eta^{\prime}\nonumber\\
\hat{n}\times\vec{B}^{a}&=&\frac{N_{f}g^{2}}{8\pi^{2}f_{\eta^{\prime}}}
			   \hat{n}\times\vec{E}^{a}\eta^{\prime}.
\eea

In the weak field approximation, using the above modified boundary conditions 
one can obtain the following relation inside the chiral bag\cite{color2} 
\beq
f_{\eta^{\prime}}^{2}m_{\eta^{\prime}}^{2}=4N_{f}\frac{g^{2}}{16\pi^{2}}
	      \frac{g^{2}}{16\pi^{2}}\langle G^{2}\rangle_{V}
\label{etaprime}
\eeq
where $\langle G^{2}\rangle_{V}$ is the amount of gluon condensate averaged
over the volume $V$.  Here one notes that in the scaling agument presented by
Witten\cite{witt} $m_{\eta^{\prime}}^{2}$ scales as $1/N_{c}$ since
$\langle G^{2}\rangle$ scales as $N_{c}^{2}$, $g^{2}$ as $1/N_{c}$ and
$f_{\eta^{\prime}}$ as $\sqrt{{N}_{c}}$.

On the other hand the pseudoscalar octet mesons inside the chiral bag obtain
their masses via the quark condensate $2\sigma=\langle \bar{u}u\rangle=\langle
\bar{d}d\rangle\cong\langle \bar{s}s\rangle$\cite{sigma}
\bea
f_{\pi}^{2}m_{\pi}^{2}&\cong&-2\sigma (m_{u}+m_{d})\nonumber\\
f_{\pi}^{2}m_{K}^{2}&\cong&-2\sigma m_{s}
\label{mpi}
\eea
and the relation (\ref{rel22}).

According to the Cheshire Cat Principle, the total masses of the pseudoscalar
octet and singlet mesons come from the contributions from the mesons both
inside (for example Eq. (\ref{etaprime}) for $\eta^{\prime}$) and outside
chiral bag.  Here for simplicity we assume that in accordance with the Cheshire Cat Principle all the masses of the mesons inside the chiral bag increase at
the same rate as the bag radius increases.  For instance let $p$ be the ratio
of the meson masses outside to those inside the chiral bag then in (\ref{a})
the numerator and denominator have the same factor $p^{2}$, which can be
cancelled out.  This leads us to the conclusion that the contributions from the
right hand side of Eq. (\ref{etaprime}) and Eq. (\ref{mpi}) should have the 
identical dependence on $p^{2}$.  In other words, the parameter $a$ should have
the same value regardless of the amount of the gluon and quark condensates.  
One can thus have the same pseudoscalar octet-singlet mixing angle insensitive
to the chiral bag radius upon the Cheshire Cat Principle.

\section{Conclusions}\par 
\setcounter{equation}{0} 
\renewcommand{\theequation}{\arabic{section}.\arabic{equation}} 

In this paper we have introduced the $\eta$ degrees of freedom outside the
chiral bag to include the physical fact that the $I=0$ member of the ground
state pseudoscalar octet mixes with the corresponding pseudoscalar singlet. 
To do this, we have modified the chiral field U to have U(3)$\times$U(3) group
structure, and also we have included the SU(3) chiral and flavor symmetry
breaking terms in the lagrangian as shown in (\ref{lagcsb}) and (\ref{lagfsb}).

From the full CBM lagrangian we have the bilinear kinetic and mass terms
(\ref{lagkin}) and (\ref{lagmass}) with matrices ${\cal M}$ and ${\cal N}$
respectively in the weak field approximation.  In the $\pi_{0}$-$\pi_{8}$
channel we have diagonalized the matrices ${\cal M}$ and ${\cal N}$ under the
transformation (\ref{trf}) to yield the $\eta$-$\eta^{\prime}$ mixing angle
$\theta=-12.7^{\rm o}$ which is in good agreement with the experimental data 
$\theta_{exp}=-10^{\rm o}\sim -23^{\rm o}$\cite{exp}.  Here one notes that
depending on what assumptions are made, the bilinear kinetic and mass terms are consistent with both $\theta=-10^{\rm o}$ and $\theta=-23^{\rm o}$ and is
unable to discriminate decisively between them\cite{exp,theta}. 

Also we have proposed the mechanism to incorporate inside the bag surface the
$\eta$-$\eta^{\prime}$ mixture consistent with the pseudoscalar octet-singlet
mixing angle by introducing the pseudoscalar composite operators originated 
from the minimal multi-quark Fock space.  In this mechanism the composite 
operators $\bar{\psi}i\gamma_{5}\lambda_{0}\psi\sim\pi_{0}$ and
$\bar{\psi}i\gamma_{5}\lambda_{8}\psi\sim\pi_{8}$ could mix to produce the 
pseudoscalar octet-singlet mixture so that one could have the same
$\eta$-$\eta^{\prime}$ mixing angle both inside and outside the chiral bag, in 
accordance with the Cheshire Cat Principle.

Finally the color anomaly has been discussed in terms of the $\eta^{\prime}$
mass  and the gluon condensate inside the chiral bag.  Moreover the
pseudoscalar octet meson masses inside the chiral bag have been described via
the quark condensate.  Assuming that the meson masses inside the chiral bag
increase at the same rate as the bag radius increases, we have obtained the
same $\eta$-$\eta^{\prime}$ mixing angle regardless of the amount of the gluon
and quark condensates, in accordance with the Cheshire Cat Principle.

\vskip 1cm 
We would like to thank Gerald E. Brown and Mannque Rho for helpful discussions.
This work is supported in part by the Korea Science and Engineering Foundation
through the CTP of SNU and by the Korea Ministry of Education under Grant No.
BSRI-96-2418.

\end{document}